\documentstyle[12pt,epsfig]{article} 
\begin{document}
\title{Flavor--oscillation clocks, continuous quantum measurements and a violation of Einstein equivalence principle.}
\author{ A. Camacho
\thanks{email: acamacho@aip.de} \\
Astrophysikalisches Institut Potsdam. \\
An der Sternwarte 16, D--14482 Potsdam, Germany.}

\date{}
\maketitle

\begin{abstract}
The relation between Einstein equivalence principle and a continuous quantum measurement 
is analyzed in the context of the recently proposed flavor--oscillation clocks, an idea pioneered by Ahluwalia and
Burgard (Gen. Rel. Grav. Errata {\bf 29}, 681 (1997)).
We will calculate the measurement outputs if a flavor--oscillation clock, which is immersed in a gravitational field, 
is subject to a continuous quantum measurement. Afterwards, resorting to the weak equivalence principle, 
we obtain the corresponding quantities in a freely falling re\-fe\-rence frame.
Finally, comparing this last result with the measurement outputs that would appear in a Minkowskian spacetime
it will be found that they do not coincide,
in other words, we have a violation of Einstein equivalence principle.
This violation appears in two different forms, namely: (i) the oscillation frequency in a freely
falling reference frame does not match with the case predicted by ge\-ne\-ral relativity, a feature previously 
obtained by Ahluwalia; (ii) the pro\-ba\-bility distribution of the measurement outputs,
obtained by an observer in a freely falling reference frame,
does not coincide with the results that would a\-ppear in the case of a Minkowskian 
spacetime. Concerning this last difference, the probability distribution differs in two directions. 
Firstly, the maximum, as function of the energy of the system (that emerges if we calculate
first the probability distribution in the original curved manifold and then, resorting to the weak
equivalence principle, we find the corresponding expression in a freely falling reference
frame) is shifted with respect to the case in which the system is in a Minkowskian spacetime.
Se\-cond\-ly, the magnitude of this maximum is not equal to the respective quantity predicted by general relativity. 
In other words, we obtain two new theoretical results that predict a violation 
of Einstein equivalence principle, and that could be measured.
\end{abstract}
\newpage
\bigskip

\section{Introduction.}
\bigskip

General relativity (GR) is one of the milestones of modern physics and many
of its theoretical predictions have already been confirmed experimentally [1].
One of the cornerstones of GR is the so called weak equivalence principle (WEP),
which is based on the principle that the ratio of the inertial mass to the
gravitational--passive--mass is the same for all bodies [2].
But in connection with this principle, we must also add that all the experimental confirmations of 
WEP have been done using classical systems, i.e., employing classical test bodies [2].
Nevertheless, in the quantum context the gravitational effects (here this phrase means the effects of a 
classical gravitational field upon a quantum particle) are not so easy to describe 
as in the classical situation. For instance, in the case of a freely falling quantum 
particle, even the definition of the time of flight probability distribution is not 
uniquely defined [3]. This last fact shows us that even the simplest concepts, 
that we may define in the interaction between classical particles and gravity, become 
unclear in the extension to quantum particles. 

We may understand the difference between the classical and quantum situations 
noting that in the former the solution to the motion equation of a particle in a gra\-vi\-tational field 
does not depend upon the mass of the involved particle, while in the latter the 
solution of the Schr\"odinger equation does depend, explicitly, upon the mass parameter [4]. 
In addition to this last fact, it is already known that the interference pattern of a thermal neutrons 
beam, moving in a homogeneous gravitational field, does depend upon the mass parameter, a fact that has already been detected ex\-pe\-rimentally [5].
This last experimental result means that gravity, at quantum level, is not a purely geometric effect [4].
Einstein equivalence principle (EEP) (in order to avert any misunderstanding here we follow
the definitions of [2], in other words, EEP means here: {\it for every pointlike event
of spacetime, there exists a sufficiently small neighborhood such that in every
local, freely falling frame in that neighborhood, all the laws of physics obey
the laws of special relativity}) could also have, at quantum level, conceptual problems, i.e., recently it has been claimed
that Feynman and Schr\"odinger formulations of quantum theory (QT) could be not equivalent in curved spacetimes [6]. 

Another very interesting point, in connection with this interplay between a cla\-ssi\-cal gravitational field 
and a quantum particle, is the consideration of a continuous quantum measurement 
in the case of a quantum particle immersed in a gravitational field. The role that
gravity and measuring processes play at quantum level, in the context of interference
experiments, shows that the mass of the corresponding test particle plays a fundamental role in the determination of the
corresponding interference pattern [7], this fact means that gravity at quantum level, even under the presence
of a measuring process, is not a purely geometric effect. The emergence of the mass
of the test particle in the corresponding interference pattern happens not only in the case of quantum demolition measurements [7], but
also in the context of the so called quantum nondemolition measurements [8],
in which a variable is measured in such a way that the complementary one does not disturb
the evolution of the chosen variable [9], i.e., the measurement outputs do not depend in this
case upon random quantum fluctuations. Concerning the appearance of the mass of the test
particle in these two different types of measuring processes, we must point out that
in the case of quantum demolition measurements the mass parameter appears always in the combination
$m/\hbar$, but it must be also added that this is not the case for a quantum nondemolition measurement. In this last situation we may find some cases
in which $m$ appears, but $\hbar$ is absent [8].

Additionally,
the possible emergence of problems in the analysis of free fall, as well as the
possible requirement of new concepts in gravitation, if a continuous quantum measurement
is introduced, has already also been analyzed [3].

To be more concrete, as Onofrio and
Viola have already pointed out [3], even the kinematical description of a freely falling
quantum particle shows conceptual difficulties. For instance, in the case of a
freely falling classical particle we may consider time of flight, but in the general quantum mechanical case we may
only introduce time--of--flight probability densities.
Nevertheless, up to now, there is no consistent definition of the time--of--flight probability density
for the case of a freely falling quantum particle. This last remark shows us that even the simplest classical
concepts do have conceptual difficulties if they are extrapolated to the corresponding quantum mechanical situation.

A second problem in this direction, once again pointed out by Onofrio and Vio\-la [3],
concerns the case in which a freely falling quantum particle is subject to a measuring process [3].
Here the difficulty emerges from the fact that in general the coupling between measuring device and measured system can
be addressed only by postulating the validity, at quantum level, of EEP. Hence we should look for a manner to
analyze the validity of EEP, and not assume it from the very begining.

An interesting idea that has rendered important results in the context of the validity of GR
at quantum realm is the concept of flavor--oscillation clocks, an idea pioneered by Ahluwalia and
Burgard [10, 11]. At this point it is noteworthy to mention that concerning these
gravitationally induced neutrino oscillation phases, some time ago, a controversy arose [12, 13, 14, 15].

This controversy was solved by Konno and Kasai [16], who proved that the arguments used in [12] were wrong.
For the sake of clarity, and also because the present work is based upon this gravitationally induced
neutrino oscillation phases, we cite Konno and Kasai tex\-tual\-ly: {\it However the authors of Ref. 4}
(T. Bhattacharya, S. Habib, and E. Mottola, Preprint gr--qc/9605074) {\it assume that di\-ffe\-rent mass eigenstates are produced at different times.
This assumption seems to be of questionable validity, because the relative phase bet\-ween the two
different mass eigenstates initially becomes arbitrary}.
Konno and Kasai have also shown that the discrepancy between the results of [10], and those appearing in [13, 14, 15]
is due to different assumptions regarding constancy along the neutrino trajectory.

The validity and correctness of the definition of these flavor--oscillation clocks
have already propeled some work and results not only within the context of GR, but also
in Brans--Dicke theory [17].

Related to a violation of EEP is the possible incompleteness of the general
relativistic description of gravity, which has been claimed in connection with 
flavor--oscillation clocks [18]. In this last work the oscillation frequencies,
of the aforementioned clocks, in a flat spacetime and in a locally inertial coordinate system (this reference system is embedded in a curved manifold)
do not coincide. In other words, the result of a local experiment in a freely falling reference frame
does not match with the result of the corresponding experiment, when it is carried out
in a flat spacetime. Clearly Ahluwalias's work implies a violation of EEP, and in order to
confront his theoretical predictions with the experiment he has introduced a proposal,
which involves the detection of oscillation frequencies [18].

In the present work we will try to find new theoretical predictions, that could allow
us to consider additional experimental proposals that could test the validity of EEP at quantum level. Therefore we will consider an idea closely related to Ahluwalia's situation [18],
namely we start with a quantum mechanical linear superposition of two different mass eigenstates, 
related to different lepton generations. This system will be immersed into a gravitational potential, 
which comprises two contributions, a constant one and a gravitational potential which has a non--vanishing gradient. 
We will then introduce a continuous quantum measurement of the energy and, using the effective Hamiltonian formalism, 
we will find the probability distribution of the possible measurement outputs
that would be obtained by an observer at rest with respect to the employed coordinate system
of the curved manifold.
Afterwards, resorting to WEP, we will deduce the corresponding results by an observer which
is in a freely falling reference frame. From these last results we will conclude that the measurement 
outputs that we would obtain in a freely falling reference frame does not match with the measurement outputs
that would emerge if the experiment were carried out in a Minkowskian spacetime, i.e.,
we deduce a violation of EEP.
 
This violation appears in two different forms, namely: (i) the oscillation frequency in a freely
falling reference frame does not match with the case predicted by ge\-ne\-ral relativity, a feature previously 
obtained by Ahluwalia; (ii) the pro\-ba\-bility distribution of the measurement outputs,
obtained by an observer in a freely falling re\-fe\-rence frame,
does not coincide with the results that would a\-ppear in the case of an observer in a Minkowskian spacetime. 
Concerning this last di\-ffe\-rence, the probability distribution differs in two directions:

(1) Firstly, the maximum, as function of the energy of the system, that appears if we calculate
first the probability distribution in the original curved manifold and then, resorting to WEP, we find the corresponding expression in a freely falling re\-fe\-rence
frame, is shifted with respect to the case in which the system is located in a Minkowskian spacetime.

(2) Se\-cond\-ly, the magnitude of this maximum is not equal to the respective quantity predicted by GR.

In other words, in this work we will obtain two new theoretical results that predict a violation
of EEP at quantum level, and that could, in principle, be measured.
\bigskip

\section{Flavor--oscillation clocks and continuous quantum measurements.}
\bigskip

Let us consider the quantum mechanical superposition of different mass eigenstates, 
for example, two neutrinos related with two different lepton generations

{\setlength\arraycolsep{2pt}\begin{eqnarray}
|\alpha; t = t_0>~= \cos(\theta)|m_1> + \sin(\theta)|m_2>,
\end{eqnarray}}

{\setlength\arraycolsep{2pt}\begin{eqnarray}
|\beta; t = t_0>~= -\sin(\theta)|m_1> + \cos(\theta)|m_2>.
\end{eqnarray}}

As was mentioned before this case has been already analyzed to study the behaviour of the
co\-rres\-ponding oscillation frequencies [18].
Clearly these two kets are orthogonal to each other.
\bigskip

In our case the background geometry is described by the following line element

{\setlength\arraycolsep{2pt}\begin{eqnarray}
ds^2 = -(1 - {2GM\over rc^2} - 2\vert\phi\vert)dt^2 + (1 + {2GM\over rc^2} + 2\vert\phi\vert)(dx^2 + dy^2 + dz^2),
\end{eqnarray}}

\noindent where $\phi$ is a non--vanishing constant term and $r$ is the distance to the center of a spherical body with 
mass $M$. We also have that $0 <|\phi| <<1$, this is a condition that has to be fulfilled in
order to have a weak field approximation [19].

This constant contribution, $\vert\phi\vert$, to the gravitational potential
could have physical meaning, for instance, it could stem from the
gra\-vi\-tational potential of the local cluster of galaxies, the so called
Great Attractor [20]. Concerning its properties in the solar system, this potential is constant
to about 1 part in $10^{11}$ [18].

At this point the consistency of (3) with Einstein's equations must be addressed. Clearly, one of our goals is to remain
always inside the context of the weak field limit of GR, indeed we have been considering
gravitational potentials, a concept that becomes unclear outside the weak field limit.
In order to comprehend better this point we may consider the affine connections
coming from (3). It is readily seen that, up to linear order in ${GM\over rc^2}$ and $\vert\phi\vert$, these connections
go like ${GM\over c^2r}(x^{i}/r)$, i.e., the term $\vert\phi\vert$ plays in the determination
of the affine connections no role at all. This last fact means that if we calculate, up to linear order in ${GM\over rc^2}$ and $\vert\phi\vert$,
Riemann, Ricci, and Einstein tensors stemming from (3), and compare them with the
corresponding tensors when the term $\vert\phi\vert$ is absent, then we find that these
tensors are the same. But the case when $\vert\phi\vert$ is absent comprises a solution to the linearized Einstein equations [19],
hence expression (3) is also a solution of these equations. In other words, the metric
that our line element defines is consistent with the linearized Einstein equations.

The idea in this work is to pursue the analysis of the joint effects of a measuring 
process and of a gravitational field, therefore in a first approach to this issue 
we will neglect the spin--dependent terms, and in consequence it will not be necessary to deal with Dirac 
equation in a curved background. 

We will also assume that our two quantum systems have vanishing small three--momentum, 
i.e., they are at rest with respect to the coordinate system defined by the metric 
implied by the line element given in (3). At this point it is noteworthy to comment that the present analysis
is carried out within the non--relativistic context, this condition has been imposed
for the sake of simplicity. Neverwithstanding, the current neutrino oscillation phenomenology
belongs to the relativistic realm [21]. This last remark means that the general case must
be also analyzed, and that the present results have to be a particular limit of the most general
situation.

Under these conditions the time evolution of our quantum superpositions
are easily evaluated along Stodolsky's ideas [22] 

{\setlength\arraycolsep{2pt}\begin{eqnarray}
|\alpha; t, t_0>~= \cos(\theta)\exp\{-{i\over\hbar}\int_{t_0}^t[\eta_{00} + {GM\over rc^2} + \vert\phi\vert~]P_1^0dt'\}|m_1>  \nonumber\\
+ \sin(\theta)\exp\{-{i\over\hbar}\int_{t_0}^t[\eta_{00} + {GM\over rc^2} + \vert\phi\vert~]P_2^0dt'\}|m_2>,
\end{eqnarray}}

{\setlength\arraycolsep{2pt}\begin{eqnarray}
|\beta; t, t_0>~= -\sin(\theta)\exp\{-{i\over\hbar}\int_{t_0}^t[\eta_{00} + {GM\over rc^2} + \vert\phi\vert~]P_1^0dt'\}|m_1>  \nonumber\\
+ \cos(\theta)\exp\{-{i\over\hbar}\int_{t_0}^t[\eta_{00} + {GM\over rc^2} + \vert\phi\vert~]P_2^0dt'\}|m_2>,
\end{eqnarray}}

\noindent where $P_1^0 = m_1c^2$, $P_2^0 = m_2c^2$.
\bigskip 

These last facts and the definitions $E_1 = (1 - {GM\over rc^2} - \vert\phi\vert)m_1c^2$ and $E_2 = (1 - {GM\over rc^2} - \vert\phi\vert)m_2c^2$, allow us to rewrite (4) and (5) as follows 

{\setlength\arraycolsep{2pt}\begin{eqnarray}
|\alpha; t, t_0>~= \cos(\theta)\exp\{{i\over\hbar}E_1(t -t_0)\}|m_1>  \nonumber\\
+ \sin(\theta)\exp\{{i\over\hbar}E_2(t -t_0)\}|m_2>,
\end{eqnarray}}

{\setlength\arraycolsep{2pt}\begin{eqnarray}
|\beta; t, t_0>~= -\sin(\theta)\exp\{{i\over\hbar}E_1(t -t_0)\}|m_1>  \nonumber\\
+ \cos(\theta)\exp\{{i\over\hbar}E_2(t -t_0)\}|m_2>.
\end{eqnarray}}

From these two last results we may now evaluate probabilities, for instance, the probability of having at time $\tilde{t}$ the system in $|\alpha; \tilde{t}, t_0>$ 
and finding it, at a time later $t$ ($\tilde{t} < t$), in $|\beta; t, t_0>$.
This probability is given by $P = |<\beta; t, t_0|\alpha; \tilde{t}, t_0>|^2$.
From the previous results we have

{\setlength\arraycolsep{2pt}\begin{eqnarray}
P = {1\over 2}\sin^2(2\theta)\Bigl[1 - \cos[{(E_2 - E_1)(\tilde{t} - t)\over\hbar}]\Bigr].
\end{eqnarray}

This is the probability calculated by an observer at rest with respect 
to the quantum systems, i.e., at rest with respect to the coordinate system defined
by expression (3).

Let us now consider a continuous quantum measurement, namely we will measure, continuously, 
the energy of the kets given by (4) and (5). 
A continuous quantum measurement can be described by means of the so called 
restricted path integral formalism [23], but in our case, in order to evaluate 
the evolution operator, it will be more useful to employ the so called
effective Hamiltonian formalism, which is equivalent to the restricted path integral
formalism [24].

According to this model, a continuous quantum measurement is described by an additional 
term in the Hamiltonian

{\setlength\arraycolsep{2pt}\begin{eqnarray}
H_{eff} = H_0 - {i\hbar\over T\Delta E^2}(H_0 - E)^2,
\end{eqnarray}

\noindent where $H_0$ is the Hamiltonian that describes the evolution without measurement, 
$\Delta E^2$ the resolution (also called the error of the measurement) of the measuring device, $T$ the time that the measuring process lasts, and $E$ the measurement output. 
Clearly, the new Hamiltonian is non--hermitian, but this is a consequence of the fact that 
in this case we have a selective measurement [24].
\bigskip

Under these new conditions, the evolution of our systems is given now by 

{\setlength\arraycolsep{2pt}\begin{eqnarray}
|\alpha; t, t_0>~= \cos(\theta)\exp\{{i\over\hbar}E_1(t -t_0) - {(E - E_1)^2\over \Delta E^2}\}|m_1>  \nonumber\\
+ \sin(\theta)\exp\{{i\over\hbar}E_2(t -t_0) - {(E - E_2)^2\over \Delta E^2}\}|m_2>,
\end{eqnarray}}

{\setlength\arraycolsep{2pt}\begin{eqnarray}
|\beta; t, t_0>~= -\sin(\theta)\exp\{{i\over\hbar}E_1(t -t_0) - {(E - E_1)^2\over \Delta E^2}\}|m_1>  \nonumber\\
+ \cos(\theta)\exp\{{i\over\hbar}E_2(t -t_0) - {(E - E_2)^2\over \Delta E^2}\}|m_2>.
\end{eqnarray}}

The probability in this new situation is

{\setlength\arraycolsep{2pt}\begin{eqnarray}
|<\beta; t, t_0|\alpha; \tilde{t}, t_0>|^2 = \nonumber\\
{1\over 4}\sin^2(2\theta)\Bigl[\exp\{-4{(E- E_1)^2\over \Delta E^2}\} \nonumber\\
+ \exp\{-4{(E- E_2)^2\over \Delta E^2}\} \nonumber\\ 
- 2 \exp\{-{2\over \Delta E^2}[(E - E_1)^2 \nonumber\\
+ (E - E_2)^2]\}\cos\{{E_1 - E_2\over\hbar}(t - \tilde{t})\}\Bigr].
\end{eqnarray}

Let us now denote $P(E) = |<\beta; t, t_0|\alpha; \tilde{t}, t_0>|^2$, clearly 
${dP\over dE} = 0$ defines the extremal values of this probability distribution, and in order to obtain the value of the 
energy that satisfies this condition we must solve a trascendental equation. 
But in order to understand how this extremal value depends upon energy let us consider
the case in which $(E - E_1)^2 <<\Delta E^2$ and $(E - E_2)^2 <<\Delta E^2$. This condition is satisfied if
we have the case in which we perform our experiment using a measuring device
which has a sufficiently large experimental error.

This last fact means also that
the measurement takes place very far from that region of the measuring process in which the
backreaction of the measuring device upon the measured system plays the leading role in the determination, for instance,
of the variance of the measurement outputs. We may rephrase this last assertion stating that in the present case
the role played by quantum noise may be neglected.
This aforementioned region, in which quantum noise becomes relevant, bears the name quantum threshold [23].

If $E = E^{\star}$ is the value of energy that satisfies the condition ${dP\over dE} = 0$,
then we have that

{\setlength\arraycolsep{2pt}\begin{eqnarray}
E^{\star} = (1 - \vert\phi\vert - {GM\over c^2r}){(m_1 + m_2)c^2\over 2}.
\end{eqnarray}

 It is also readily calculated that (under our approximation)

{\setlength\arraycolsep{2pt}\begin{eqnarray}
{dP^2\over dE^2} = -{4\over\Delta E^2}\sin^2{(2\theta)}\Bigl[1 - \cos{\Bigl({(E_1 - E_2)(t -\tilde{t})\over\hbar}\Bigr)}\Bigr].
\end{eqnarray}

Expression (14) implies that condition ${dP\over dE} = 0$ is related to a maximum, and not to a minimum.

From the last results we have now that the maximum of this probability is

{\setlength\arraycolsep{2pt}\begin{eqnarray}
P(E = E^{\star}) = {1\over 2}\sin^2(2\theta)
\Bigl[1 - {4\over\Delta E^2}[1 - \vert\phi\vert - {GM\over c^2r}]^2[m_1 - m_2]^2c^4\Bigr]\nonumber\\
\times\Bigl[1 - \cos[(1 - \vert\phi\vert - {GM\over c^2r}){(m_1 - m_2)c^2(t - \tilde{t})\over \hbar}]\Bigr].
\end{eqnarray}

At this point we must remember that from the very begining it has been assumed that
our quantum particles are at rest with respect the coordinate system under consideration.
Expression (15) is the maximum of the probability of the measurement outputs detected
by an observer at rest with respect to the coordinate system defined by expression (3), which means that he is also at rest with respect
to $m_1$ and $m_2$.

In order to analyze the predictions of GR we resort now to WEP, i.e., we consider 
now a locally inertial frame. The condition of vanishing three--momentum means that
we must choose that locally inertial frame in which the oscillation clock is momentarily
at rest.

As has already been pointed out [18], WEP allows
us to annul the gradients of the gravitational potential, but it can not discard 
its constant parts. This remark means that if we resort to WEP in expression (15), 
then we may annul ${GM\over c^2r}$ but $\vert\phi\vert$ remains, and in consequence 
the maximum probability, that an observer in a freely falling coordinate system 
detects, is 
 
{\setlength\arraycolsep{2pt}\begin{eqnarray}
P_F(E = E^{\ast}) = {1\over 2}\sin^2(2\theta)
\Bigl[1 - {4\over\Delta E^2}[1 - \vert\phi\vert~]^2[m_1 - m_2]^2c^4\Bigr]\nonumber\\
\times\Bigl[1 - \cos[(1 - \vert\phi\vert){(m_1 - m_2)c^2(t - \tilde{t})\over \hbar}]\Bigr].
\end{eqnarray}

This probability appears if the energy has the value

{\setlength\arraycolsep{2pt}\begin{eqnarray}
E^{\ast} = (1 - \vert\phi\vert){(m_1 + m_2)c^2\over 2}.
\end{eqnarray}

Nevertheless, the predictions stemming from EEP (the measurement outputs
if the experiment were carried out in a Minkowskian spacetime) read

{\setlength\arraycolsep{2pt}\begin{eqnarray}
P_M(E = E^{\dagger}) = {1\over 2}\sin^2(2\theta)
\Bigl[1 - {4\over\Delta E^2}[m_1 - m_2]^2c^4\Bigr]\nonumber\\
\times\Bigl[1 - \cos[{(m_1 - m_2)c^2(t - \tilde{t})\over \hbar}]\Bigr],
\end{eqnarray}

{\setlength\arraycolsep{2pt}\begin{eqnarray}
E^{\dagger} = {(m_1 + m_2)c^2\over 2}.
\end{eqnarray}

This last energy is the value that renders, in this case, the maximum of the probability distribution,
once again we have the approximation of a very large experimental error.

Clearly, the measurement outputs, with respect to a freely falling reference frame, 
contain the information concerning the constant parts of the gravitational potential, 
a consequence of WEP, which leads to a violation of EEP. Indeed, $P_M$ does not 
coincide with $P_F$, i.e., the result in a freely falling reference frame does 
not match with the result in a flat spacetime. Resorting to WEP leads us to a violation 
of EEP.

This violation of EEP comprises three aspects. The first one has already been pointed out [18], 
namely the frequencies are not the same. This is readily seen if we compare the arguments of
the cosine functions in expressions (16) and (18).

But now we have two new aspects that define two differences, which in principle could be detected:

(1) Firstly, the maximum of the probability (as a function of the energy of our quantum system) is shifted. 
According to EEP this maximum is related to the fo\-llow\-ing value of the energy
$E^{\dagger}= {(m_1 + m_2)c^2\over 2}$. If we evaluate first everything in our original curved manifold,
and afterwards we resort to WEP (which means that we are now at rest with respect to a freely
falling reference frame), then the maximum of
the probability emerges if the energy has the value $E^{\ast} = (1 - \vert\phi\vert){(m_1 + m_2)c^2\over 2}$.
Then we deduce that $E^{\ast} < E^{\dagger}$, in other words, the curve of the
probability distribution is, in the case of a freely falling observer, shifted to a zone of smaller energy values, compared with the corresponding curve of an
observer in a Minkowskian spacetime. This shift comprises a violation of EEP, and defines
an experimental proposal that could allow us to confront the validity, at quantum level, of GR.

(2) Secondly, the maximum is not the same,
$P_F/P_M \not = 1$. Once again this can be seen from expressions (16) and (18).

We have a new local experiment that allows us
to determine if we are in a region in which there is a non--vanishing gravitational potential.
In other words, resorting to WEP our construction leads to a violation of EEP.
\bigskip

\section{Conclusions.}
\bigskip

Employing a flavor--oscillation clock, defined with the superposition of different mass eigenstates of two neutrinos related with different lepton generations,
we found the measurement outputs of a selective continuous quantum measurement of
the energy of our oscillation clock. The whole system was at rest with respect
to an observer embedded in gravitational field, whose potential comprised two
parts, namely a gradientless contribution and a term coming from a spherically symmetric body
with mass $M$. Afterwards, by means of WEP, the measurement outputs in a freely falling coordinate system
were found. The maximum probabilities, and the energies at which these maximums
emerge, where evaluated (under the condition that the measuring device performs
a very rough measurement), and it was found that the quantitites in a freely
falling reference frame are not the same that EEP predicts. In other words,
we have a local experiment in which the predictions stemming from EEP do not
coin\-cide with the results that we obtain if we evaluate our results first in
our initial curved manifold and after this, resorting to WEP, we find the measurement outputs
with respect to a locally inertial reference frame.

Clearly the present results
do contain Ahluwalia's conclusions [18], and in this sense our work is a generalization
of Ahluwalia's case. Indeed, here we predict not only Ahluwalia's violation of EEP (the frequencies of the two involved flavor--oscillation clocks do not coincide),
but we also add two new experimental facts (the shift of the probability curve and the modification of the magnitude of the maximum of this curve) that, in principle, could be detected, and that
lead also to a violation of EEP.

This violation of EEP stems from the fact that in QT the evolution
operator, even in the case of a continuous quantum measurement, involves gravitational
potentials and not derivatives of them. A further consequence of our result comes
from the fact that EEP contains, implicitly, two assumptions,
namely (i) local Lorentz invariance and (ii) local position invariance [2]. Taking a look
at the fact that this violation of EEP appears as a consequence of the possibility
of detecting constant terms in the gravitational potential, expressions (15) and (16),
then we conclude that two regions, having two different constant gravitational potentials,
$\vert\phi_1\vert$ and $\vert\phi_2\vert$, would determine different measurement outputs, i.e., the result of an experiment depends upon the region in which
the measurement takes place. In other
words, this EEP violation could imply also the violation of this aforementioned local position invariance.
The possible violation of this property was first noted by Ahluwalia [25].

Experimentally this violation could be tested from the fact that the probability distribution curve
of a freely falling observer is shifted to a zone with smaller energy values, compared with the corresponding curve of an
observer in a Minkowskian spacetime. This shift comprises a violation of EEP, and defines
an experimental proposal that could allow us to confront the validity, at quantum level, of some of the postulates behind GR.
In particular the topic of the validity of the local position invariance postulate could
be analyzed.

An additional issue, which must be also addressed, concerns the feasibility
of these kind of experiments. The possibility of having different constant terms could be achieved using the
case of a hollow cylinder filled with mercury [18], an experimental proposal which seems to be not very far
from the present technological capabilities. An additional experimental advantage of our proposal
lies in the fact that we do not need an experimental device with a very small
experimental error, in other words, it is not ne\-ce\-ssary to perform the experiment very near
the quantum threshold of our oscillation clock [23].
This can be understood remembering the approximation done after expression (12),
which means that we do have a measuring device with a large experimental error.

The independent physical significance, at quantum realm, of the gravitational potentials,
could mean not only the breakdown of EEP but, as has also already been pointed out [26], it could imply the appearance of
non--locality in quantum theory. This issue is still a controversial point in
the context of QT [27]. In other words, at quantum level,
the role that the elements of the metric could play would imply not only the violation of EEP and of local
position invariance, but also the emergence of non--locality.

As Konno and Kasai have already mentioned (see section 5 of [16]), the cu\-rrent
technology has problems to provide verification of the role that quantum effects and gravitational effects could play, simultaneously,
in some systems (in their remark they do not include the experimental status
in the context of quantum measurement). Nevertheless, progress in technology may make the experimental
verification of such effects po\-ssi\-ble, and this fact provides an additional reason
for further investigation in this topic.
Concerning the experimental status in the context of quantum measurement, it has to be mentioned that nowadays the kind of experiments that demands a continuous quantum measurement
are outside the current technological possibilities. Neverwithstanding, it is also important to add that
they could be feasible in the future [28]. From the last arguments we may assert that the present experimental proposal
could shed some light not only on the old conundrum of the quantum measurement problem, but also on the validity of EEP at quantum level.

In this relation between measured system and measuring device we may find some other problems that remain unsolved, for
instance, the question around the validity of EEP in the coupling between measuring device and test
mass in the case of a freely falling quantum particle which is subject to a continuous measurement [3].

\bigskip

\Large{\bf Acknowledgments.}\normalsize
\bigskip

The author would like to thank A. A. Cuevas--Sosa and A. Camacho--Galv\'an for their 
help, and D.-E. Liebscher for the fruitful discussions on the subject. 
The hospitality of the Astrophysikalisches Institut Potsdam is also kindly acknowledged. 
This work was supported by CONACYT Posdoctoral Grant No. 983023.

\end{document}